\documentclass[11pt,floatfix,amssymb,prx,aps,longbibliography,superscriptaddress,nopacs]{revtex4-2}

\usepackage{comment}

\newcommand{\om}{\left( \omega \right)}

\newcommand{\Vm}{V^{\scriptscriptstyle{\mathrm{M}}}}

\newcommand{\nX}{n_{\scriptstyle{x}}}
\newcommand{\nY}{n_{\scriptstyle{y}}}

\newcommand{\OmegaX}{\Omega_{\scriptstyle{x}}}
\newcommand{\OmegaY}{\Omega_{\scriptstyle{y}}}
\newcommand{\OmegaZ}{\Omega_{\scriptstyle{z}}}

\newcommand{\GY}{\Gamma_{\scriptstyle{y}}}
\newcommand{\GX}{\Gamma_{\scriptstyle{x}}}

\newcommand{\gX}{g_{\scriptstyle{x}}}
\newcommand{\gY}{g_{\scriptstyle{y}}}
\newcommand{\gb}{g_{\mathrm{b}}}

\newcommand{\Omegamj}{\Omega_{j}}

\newcommand{\xX}{x}
\newcommand{\xY}{y}
\newcommand{\pX}{p_{\scriptstyle{x}}}
\newcommand{\pY}{p_{\scriptstyle{y}}}

\newcommand{\supp}{s}
\newcommand{\splitt}{\delta}

\usepackage{amsmath}
\usepackage[dvips]{graphicx}
\usepackage{epsfig}
\usepackage{tabularx}
\usepackage{lineno}
\usepackage{ulem}
\usepackage{xcolor}
\usepackage{soul}

\begin{document}

\title{High purity two-dimensional levitated mechanical oscillator}

\author{Q. Deplano}
\email{These authors contributed equally}
\affiliation{Dipartimento di Fisica e Astronomia, Universit\`a di Firenze, via Sansone 1, I-50019 Sesto Fiorentino (FI), Italy}
\affiliation{INFN, Sezione di Firenze, via Sansone 1, I-50019 Sesto Fiorentino (FI), Italy}

\author{A. Pontin}
\email{These authors contributed equally}
\affiliation{CNR-INO, largo Enrico Fermi 6, I-50125 Firenze, Italy}

\author{A. Ranfagni}
\affiliation{Dipartimento di Fisica e Astronomia, Universit\`a di Firenze, via Sansone 1, I-50019 Sesto Fiorentino (FI), Italy}

\author{F. Marino}
\affiliation{CNR-INO, largo Enrico Fermi 6, I-50125 Firenze, Italy}
\affiliation{INFN, Sezione di Firenze, via Sansone 1, I-50019 Sesto Fiorentino (FI), Italy}

\author{F. Marin}
\email[Electronic mail: ]{francesco.marin@unifi.it}
\affiliation{Dipartimento di Fisica e Astronomia, Universit\`a di Firenze, via Sansone 1, I-50019 Sesto Fiorentino (FI), Italy}
\affiliation{European Laboratory for Non-Linear Spectroscopy (LENS), via Carrara 1, I-50019 Sesto Fiorentino (FI), Italy}
\affiliation{INFN, Sezione di Firenze, via Sansone 1, I-50019 Sesto Fiorentino (FI), Italy}
\affiliation{CNR-INO, largo Enrico Fermi 6, I-50125 Firenze, Italy}


\begin{abstract}
In recent years, levitated optomechanics has delivered on the promise of reaching the motional quantum ground state. An important next milestone of the field would be the generation of mechanical entanglement. An ideal candidate is the two-dimensional motion in the polarization plane of an optical tweezer inside an optical cavity, where optical and mechanical modes are coupled via coherent scattering. Achieving this goal requires two key conditions: two-dimensional ground state cooling along with substantial spectral overlap between the two modes. The latter is essential to generate the necessary correlations, but unfortunately, it hinders efficient cooling thus narrowing the useful parameter space. In this work, we report the achievement of a high purity two-dimensional state in a regime where the strong optomechanical coupling induces the desired spectral overlap between oscillations in different directions, as reflected in the non-trivial spectral shape of the detected cavity field. As a result, significant correlations consistently arise between any pair of orthogonal directions, preventing the motion from being reduced to two independent one-dimensional oscillators and leading to higher purity compared to that scenario. Our system serves as an excellent platform for realizing continuous variable entanglement in two-dimensional motion.

\end{abstract}

\maketitle

\section*{Introduction}
Levitated optomechanical systems provide a powerful platform for the manipulation of mesoscopic quantum objects with applications ranging from fundamental physics~\cite{Li2010,Moore2022,Gratta2014} to quantum sensing~\cite{Geraci2010,Tania2024} and technologies~\cite{Vinante2024}. Some of these systems have been cooled near the zero-point energy~\cite{Delic2020,Ranfagni2022,Piotrowski2023,Magrini2021,Tebbenjohanns2021} opening the way towards more refined quantum experiments including the preparation of novel quantum states~\cite{RomeroIsart2024,Neumeier2024,Nothup2024,Frimmer2024} and tests of the quantised nature of gravity~\cite{Bose2017}.

The motion of a levitated nanoparticle in the transverse plane of a tightly focused laser beam (optical tweezer \cite{Ashkin1970}) in high vacuum \cite{Millen2020,Gonzalez-Ballestero2021} offers a valuable opportunity to realise a two-dimensional oscillator with quantum properties. The optical potential generated by the tweezer light is proportional to its intensity, whose profile has an elliptical shape near the focus. In the transverse plane it is well approximated by a paraboloid that defines two orthogonal axes to which different natural frequencies of the oscillatory motion of the nanoparticle are associated. We will call them $X$ and $Y$ axes. The axis corresponding to the tighter focusing direction ($X$), providing the highest oscillation frequency, is typically orthogonal to the main polarization axis of the tweezer \cite{Novotny2012}. 

By placing the levitated particle inside an optical cavity with a suitable resonance frequency, a mode of the cavity field is populated by the scattered tweezer light. The oscillatory motion of the nanoparticle is coupled to the cavity field via this coherent scattering \cite{Vuletic2000,Windey2019,Delic2019A,Quidant2020}. When the particle is positioned on a node of the cavity standing wave, it is precisely the motion along the cavity axis ({\it bright mode}) that is coupled to the cavity field \cite{Toros2020B,Toros2021,Borkje2023}. Consequently, the $X$ and $Y$ oscillations have optomechanical coupling rates proportional to $\sin \theta$ and $\cos \theta$ respectively, where $\theta$ is the angle between the $Y$ direction and the cavity axis, which is assumed to be orthogonal to the tweezer axis. If $\theta$ is close to $90^{\circ}$, the $X$ oscillation can be optically cooled very efficiently by red detuning the tweezer light with respect to the cavity resonance. Thermal occupancies below unity \cite{Delic2020} (as low as 0.5 \cite{Ranfagni2022,Piotrowski2023}) have actually been achieved. On the other hand, an angle $\theta$ close to $45^{\circ}$ allows to obtain significant optomechanical coupling and cooling for both the $X$ and the $Y$ motion. If the two mechanical resonances remain well separated with respect to their width, which is enhanced by the optomechanical damping, the planar motion can still be considered as the sum of two independent mechanical modes, which could be jointly cooled near \cite{Ranfagni2022} and even both below \cite{Piotrowski2023} unity occupation number. 

If the two eigenfrequencies are close to each other, the full potential of a two-dimensional quantum system emerges thanks to the spectral overlap of the $X$ and $Y$ modes. For instance, it enabled the observation of vectorial polaritons \cite{Ranfagni2021} and the cancellation of the quantum backaction \cite{Ranfagni2023}. On the other hand, two-dimensional cooling becomes more difficult because the motion orthogonal to the cavity axis ({\it dark mode}) is not directly coupled to the optical field, but it is simply sympathetically cooled by the bright mode \cite{Toros2021} with a rate proportional to the difference between the two eigenfrequencies.   

In this work, we push the optical cooling of the two-dimensional motion of a levitated nanosphere close to the ground state (i.e., achieving thermal occupancy well below unity in all directions) by going beyond the optomechanical weak-coupling regime. By maintaining significant spectral overlap between the $X$ and $Y$ modes we ensure that they both largely share the same bath, in which quantum fluctuations play a major role. 

The resulting correlations are a fundamental characteristic of our two-dimensional dynamics, which cannot be simply decomposed into the sum of independent orthogonal oscillations.

As a first indicator of quantumness, we calculate the global state purity. We show that, thanks to the correlation between $X$ and $Y$, its value is significantly greater than the simple $ 1/(2\nX + 1)(2 \nY +1) $ attained by independent oscillators having the thermal occupancies $\nX$ and $\nY$. This indicates the enhanced quantumness of the system. To quantify this further, we evaluate the quantum discord, i.e., the quantum component of the mutual information between the two oscillators, and show that it is indeed significantly greater than zero. The two high purity oscillators then also exhibit quantum correlations and thus provide an important platform for applications in quantum information and sensing.

\section*{Experiment}

\begin{figure}[!htb]
    \centering
    \includegraphics[scale=0.5]{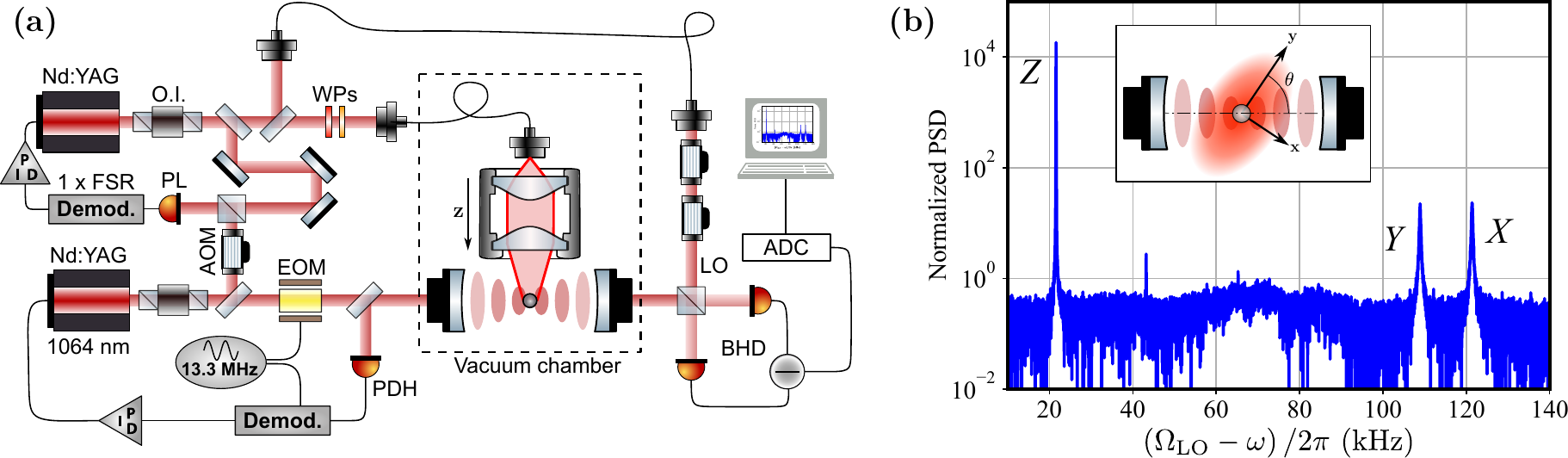}
    \caption[Fig1]{\textbf{Overview of the experiment}. a) Simplified scheme of the experimental setup. OI: optic isolator, WP: wave plate, PL: phase locking photodiode, AOM: acousto-optic modulator, EOM: electro-optic modulator, PDH: Pound-Drever-Hall detection, LO: local oscillator, BHD: balanced heterodyne detection, ADC: analog-to-digital converter. b) Power spectral density (PSD) of the heterodyne signal for a detuning $\Delta/2\pi = -250\,$kHz. We show the anti-Stokes sideband, and we report in the abscissa $(\Omega_{\mathrm{LO}}-\omega)/2\pi$. The spectrum is normalized to the measured shot noise which is then subtracted (the dark noise, which is $\sim10\,$dB lower than the shot noise, is preliminarily subtracted from all spectra).
    The three resonance peaks corresponding to the $X$, $Y$ and $Z$ modes are identified. The inset shows the scheme of a plane orthogonal to the tweezer axis, where $Y$ denotes the tweezer polarization axis, and $X$ its orthogonal direction.}
    \label{Fig1}
\end{figure}

A $100\,$nm silica nanosphere is loaded onto an optical tweezer in a first chamber under low vacuum conditions, then transferred to a second tweezer in the science chamber at a pressure of about $10\,$mbar \cite{Calamai2021}. The science tweezer is based on $250\,$mW light power generated by a Nd:YAG NPRO laser at $1064\,$nm. A doublet of aspheric lenses, with focal lengths of $18.4\,$mm and $3.1\,$mm respectively, collects the light from a polarization maintaining fiber and refocuses it with a waist narrower than $1\,\mu$m. This optical system can be positioned with nanometric precision inside an optical cavity whose optical axis is orthogonal (within $\sim 1^{\circ}$) to the axis of the tweezer. The light is linearly polarized along a direction at $\sim 45^{\circ}$ with respect to the cavity axis (Fig. \ref{Fig1}). The oscillation frequencies of the nanosphere in the optical potential are, respectively, $(\OmegaX, \OmegaY, \OmegaZ)/2\pi = (121.1,108.5,21.4)\,$kHz for the $(X, Y, Z)$ axes.

The optical cavity has a linewidth of $\kappa/2\pi = 57\,$kHz (full width at half maximum) and it is made with a pair of equal concave mirrors in a nearly concentric configuration, giving a free-spectral-range of $\mathrm{FSR} = 3.07\,$GHz. An auxiliary Nd:YAG
laser is frequency-locked to the optical cavity, while the tweezer laser is phase-locked to the auxiliary laser with a tunable frequency offset equal to $\mathrm{FSR} + \Delta/2\pi$. This setup precisely determines the detuning $\Delta$ of the tweezer radiation from a cavity resonance. 
The light scattered into the cavity mode and transmitted through the end mirror is analyzed using a balanced heterodyne detection. 

After the transfer, the tweezer light is red detuned with respect to a cavity resonance, the nanoparticle is positioned on the cavity axis in correspondence of a node of the standing wave, and the science chamber is pumped down to a pressure of about $3\times10^{-8}\,$mbar. 

The spectrum of the heterodyne signal, normalized to shot noise, can be expressed as
\begin{equation}
S_{\mathrm{out}}(\Omega_{\scriptscriptstyle{\mathrm{LO}}}+\omega) 
= 1 + \eta \,\kappa \,\left|\chi_\mathrm{c}\left(\omega\right)\right|^2 \,\gb^2 \,S_{x_\mathrm{b} x_\mathrm{b}}\om
\label{eq_Sout}
\end{equation}
where 
$\Omega_{\scriptscriptstyle{\mathrm{LO}}}$ is the angular frequency of the local oscillator (in our experiment, we set $\Omega_{\scriptscriptstyle{\mathrm{LO}}}/2\pi = 900\,$kHz using two consecutive AOMs, working on opposite orders to blue-shift the local oscillator beam), $\eta$ is the overall detection efficiency, and $\gb$ is the optomechanical coupling rate for the motion along the cavity axis. 
The displacement spectrum $S_{x_\mathrm{b} x_\mathrm{b}}$ of the bright mode appears filtered by the optical susceptibility $\chi_\mathrm{c}(\omega)=\left[-i(\Delta+\omega)+ (\kappa/2)\right]^{-1}$. It can be written as \cite{SM1}:
\begin{equation}
S_{x_\mathrm{b} x_\mathrm{b}}\om=\frac{4}{\gb^2}\,
\frac{\gX^2 \,\GX \left|\chi_x(\omega)\right|^2
+\gY^2 \,\GY\left|\chi_y(\omega)\right|^2+\left|\gX^2 \chi_x(\omega)+\gY^2 \chi_y(\omega)\right|^2
\kappa \left|\chi_\mathrm{c}\left(-\omega\right)\right|^2}
{\left|1-2 i \chi_\mathrm{c}^{-}\left(\gX^2 \chi_x(\omega)+\gY^2 \chi_y(\omega)\right)\right|^2}
\label{eq:Sxb}
\end{equation}
where we have defined $\chi_{c}^{-}=\chi_{c}\om - \chi_{c}^*\left(-\omega\right)$ and the mechanical susceptibilities are $\chi_{j}\om=\Omegamj \left[\Omegamj^2-\omega^2-i \gamma_{j} \omega\right]^{-1}$ (with $j = x, y$). $\gamma_j$, $g_j$ and $\Gamma_j$ are respectively the rates of gas damping, optomechanical coupling, and decoherence.

In Fig. \ref{Fig1}b we display an example of such a spectrum, acquired for a detuning of $\Delta/2\pi = -250\,$kHz. The oscillations along the $X$ and $Y$ axes, projected along the cavity axis, produce two clear peaks, broadened and shifted by the optomechanical coupling.
With the detuning closer to the mechanical frequencies, both modes are cooled and broadened more effectively, so that their spectra largely overlap. This is clearly visible in the heterodyne spectrum displayed in Fig. \ref{Fig2}, which is acquired at a detuning of $\Delta/2\pi = -111\,$kHz. 

\section*{Discussion}

\begin{figure}[!htb]
    \centering
      \includegraphics[scale=0.65]{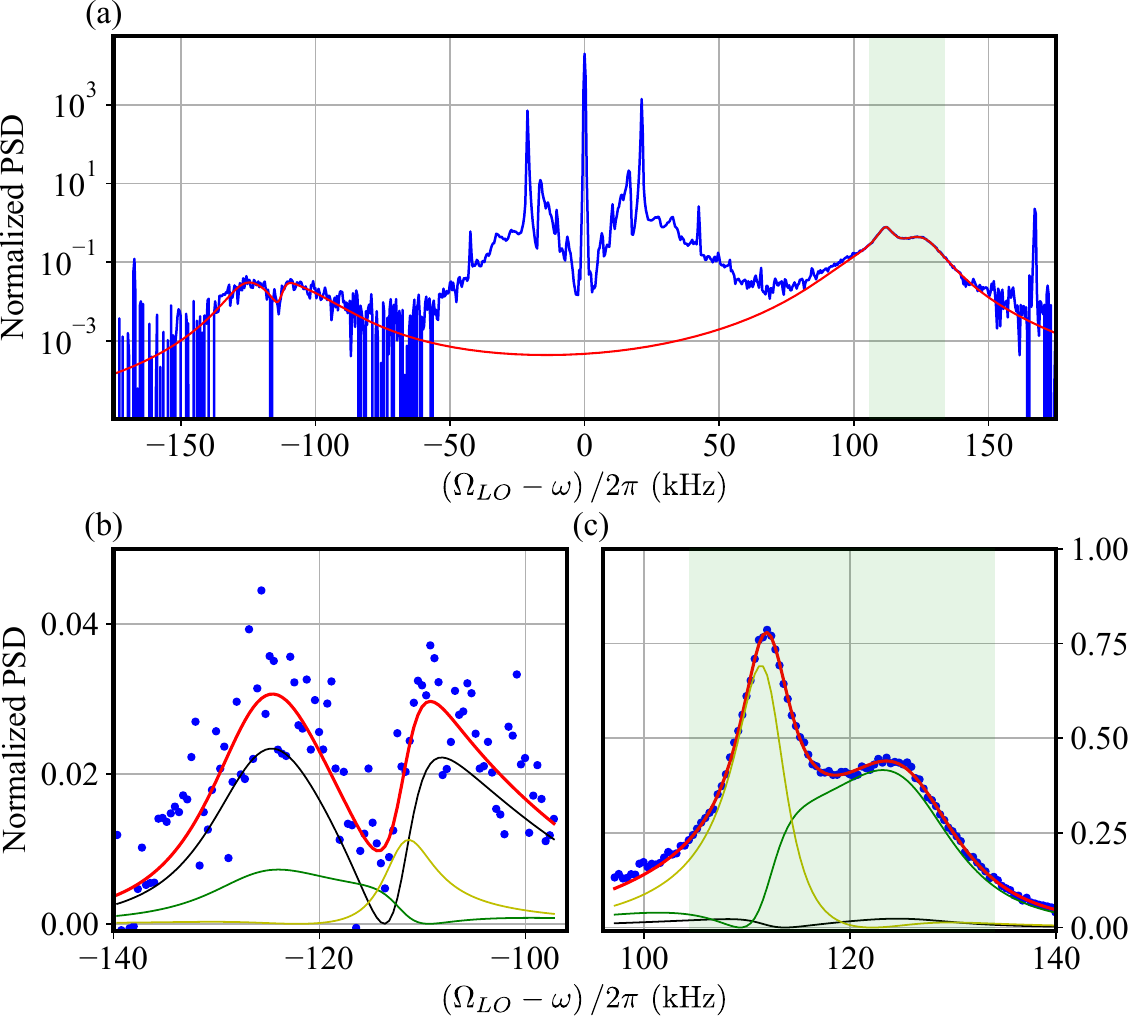}
    \caption[Fig2]{\textbf{Power Spectral Density of the heterodyne signal}. The spectrum is normalized to the measured shot noise which is then subtracted (the dark noise is preliminarily subtracted from all spectra). The abscissa is the frequency difference with respect to the local oscillator. The detuning is $\Delta/2\pi = -111\,$kHz. The red solid line shows the fit of Eqs. (\ref{eq_Sout}-\ref{eq:Sxb}) to the experimental anti-Stokes sideband (the spectral region used for the fit is shaded). The lower panels display enlarged views of the left (b) and right (c) sidebands, where different contributions to the fitted curves are shown in dark green (term $\propto \GX$), yellow (term $\propto \GY$), and black (quantum noise, term proportional to $\kappa$).}
    \label{Fig2}
\end{figure}

In Fig. \ref{Fig2}a we show that the model yielding Eqs. (\ref{eq_Sout}-\ref{eq:Sxb}) well fits the experimental data in a wide frequency range where the heterodyne spectrum is dominated by the motion in the $X-Y$ plane. Outside this region, the narrow peak given by the much warmer $Z$ motion is clearly visible around $21\,$kHz, as well as broader structures due to erratic frequencies of the rotational motion~\cite{Barker2023,Ulbricht2018}. The theoretical curve fitted to the right (anti-Stokes) sideband well reproduces also the weaker left sideband. This confirms the validity of the independently measured parameters, in particular the detection efficiency $\eta$ which plays a crucial role in our evaluation of the decoherence rates and consequently of the thermal occupancies and state purity.   

In the numerator of Eq. (\ref{eq:Sxb}), we can identify the contributions of the classical noise sources acting on the $X$ and $Y$ oscillators, quantified by the decoherence rates $\Gamma_j$, and of the quantum bath provided by
the optical vacuum noise. In Figs. \ref{Fig2}b-c we highlight that all three noise sources yield relevant and distinct contributions to the spectral shape. It is therefore clear that measuring the projection on the cavity axis is sufficient to fully characterize the two-dimensional motion. Its relevant parameters, deduced from the fit, are reported in Table \ref{tab_parametri}. 


\begin{table}
\begin{tabular}{|c|c|c|c|c|c|}
\hline
  $\OmegaX/2\pi$ (Hz)  &  
   $\OmegaY/2\pi$ (Hz)   & 
 $\gX/2\pi$ (Hz)    & 
   $\gY/2\pi$ (Hz)   &  
 $\GX/2\pi$ (Hz)    &  
  $\GY/2\pi$ (Hz)     \\
  \hline
  $\,122170\pm120\,$     &  $\,109370\pm150\,$    & $\,14130\pm220\,$    &  $\,10370\pm160\,$   &  $\,4030\pm200\pm120\,$   & $\,3050\pm170\pm90\,\,$  \\
    \hline
\end{tabular}
\caption{Parameters extracted from the fit of of the heterodyne spectrum (left sideband) with the model of Eqs. (\ref{eq_Sout}-\ref{eq:Sxb}). The reported uncertainty is the spread (one standard deviation) on five independent acquisitions. For the decoherence rate, the second quoted error is due to an uncertainty of $5\%$ in the detection efficiency, whose independent measurement yields $\eta = 0.32$.}
\label{tab_parametri}
\end{table}

We remark that the spectral contributions due to classical noise are frequency symmetric in $S_{x_\mathrm{b} x_\mathrm{b}}$, and differ in the two sidebands of $S_{\mathrm{out}}$ only due to cavity filtering. In contrast, quantum noise is present almost exclusively in the Stokes motional sideband where, in our experiment, it largely overwhelms classical noise. This produces markedly different shapes in the two sidebands, a feature that is a clear signature of quantum two-dimensional motion \cite{Ranfagni2023} and testifies to the low effective temperature achieved.

The quantum steady state of the optomechanical system is characterized by its covariance matrix $V_{ij} = 0.5\langle \{u_i , u_j\} \rangle$ where $u^T = (Q,\,  P,\, \xX,\, \pX,\, \xY,\, \pY)$, $P$ and $Q$ are the two quadratures of the intracavity field, $\xX$ and $\xY$ are the positions and $\pX$ and $\pY$ the momenta of the $X$ and $Y$ oscillators normalized to the respective zero-point fluctuations. The covariance matrix can be calculated using the Lyapunov equation $\,AV+VA^T = -D$ with the drift matrix 
\begin{equation}
    A=
    \left(
    \begin{array}{cccccc}
        -\kappa/2 & -\Delta & 0  & 0 & 0  & 0 \\
         \Delta & -\kappa/2 & 2 \gX  & 0 & 2 \gY  & 0 \\
         0 & 0 & 0  & \OmegaX & 0  & 0 \\
         2 \gX & 0 & -\OmegaX  & -\gamma_x & 0  & 0 \\
         0 & 0 & 0  & 0 & 0  & \OmegaY \\
         2 \gY & 0 & 0  & 0 & -\OmegaY  & -\gamma_y \\
    \end{array}
    \right)
\end{equation}
and the diffusion matrix $D = \mathrm{Diag} [\kappa,\,\kappa,\,0,\,4\GX,\,0,\,4\GY]$. In particular, the two-dimensional motion is described by the covariance matrix of the mechanical system $\Vm$, formed by the last four rows and columns of $V$. 

The covariance matrix $\Vm$ calculated for our system, with the parameters of Table \ref{tab_parametri}, is the following
\begin{equation}
\Vm = 
    \left(
    \begin{array}{cccc}
         2.13 & 0 & -0.32 & -0.59  \\
         0 & 2.07 & 0.52 & -0.34  \\
         -0.32 & 0.52 & 2.47 & 0  \\
         -0.59  & -0.34 & 0 & 2.48 
    \end{array}
    \right)  \, .
\end{equation}
The thermal occupancy $\nX$ of the $X$ mode, considered as a one-dimensional oscillator, is given by $\left(2 \nX+1\right) = \sqrt{\langle \xX^2 \rangle \, \langle \pX^2 \rangle}\,$ \cite{Borkje2023}, and an equivalent relation holds for the $Y$ mode. Therefore, the diagonal of $\Vm$ allows us to infer the steady-state occupancies of both modes. We derive $\nX = 0.55\pm 0.03$ and $\nY = 0.74 \pm 0.04$, where the error considers the statistical fluctuations between different measurements, as well as the systematic uncertainty in the detection efficiency. Both figures are well below unity, a threshold traditionally considered in optomechanics, indicating that the ground state of each one-dimensional oscillator is occupied with probability exceeding $50\%$. The actual probability that the system is in its two-dimensional ground state is calculated in the Supplementary Information.

However, the two thermal occupancies are not sufficient to characterize the two-dimensional motion. The 2x2 off-diagonal blocks of the covariance matrix $\Vm$, containing the correlation terms between the two oscillators, are indeed relevant in our system. A more appropriate parameter for quantifying the quantum character of the two-dimensional state is its purity, defined as
$\mu = \mathrm{Tr (\hat{\rho}^2})$
where $\hat{\rho}$ is the density matrix representing the state. It can be evaluated as the inverse square root of the determinant of the covariance matrix $\Vm$ \cite{Serafini2004,Borkje2023}. For our system, we obtain $\mu = 0.209\pm0.013$, higher than the value of $1/(2 \nX+1)(2 \nY+1) = 0.192$ that would have been derived in the case of independent oscillators with the same thermal occupancy. 
The difference between the observed purity and that
estimated for independent thermal oscillators is more
robust to fluctuations in the system parameters, and it is $0.0167\pm0.0003$.

\begin{figure}[!htb]
    \centering
      \includegraphics[scale=1.0]{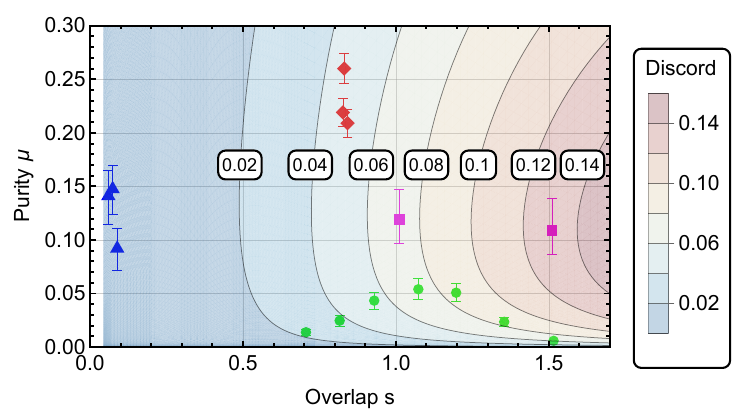}
    \caption[Fig3]{\textbf{Two-dimensional state purity $\mu$ for the motion of a nanoparticle in the tweezer transverse plane.} We show the results obtained in different experiments, as a function of the overlap parameter $\supp = 2 (\gX^2+\gY^2)/\kappa \delta$. Green dots: Ref. \cite{Ranfagni2022}. Magenta squares: Ref. \cite{Ranfagni2023}. Blue triangles: Ref. \cite{Piotrowski2023}. Red diamonds: this work, including the additional data sets \cite{SM1}. In order to provide an indication of the two-dimensional correlations present in the mechanical system, we also show the contour plot of the symmetrized quantum discord $0.5 (\mathcal{D}_{X \leftarrow Y} + \mathcal{D}_{Y \leftarrow X})$, calculated with the following parameters: $\gX/2\pi = \gY/2\pi = 12400$ Hz,  $\GX/2\pi = \GY/2\pi$ varying between 100 Hz and 300 kHz, $\Delta = -0.5(\OmegaX+\OmegaY)$. The other parameters are chosen similar to those of the present work for $\supp >  0.7$, while for $\supp < 0.7$ they change to keep realistic ranges. In details, for $\supp > 0.7$ we use $\kappa/2\pi = 57$ kHz and $(\OmegaX+\OmegaY)/4\pi = 116$ kHz. For $\supp < 0.7$ both the cavity width and the mean oscillation frequency increase, reaching $\kappa/2\pi = 330$ kHz and $(\OmegaX+\OmegaY)/4\pi = 246$ kHz when $\supp = 0.07$, thus approaching the parameters of \cite{Piotrowski2023}. The variations laws are $\kappa/2\pi = [57 + (330-57)x^4]\,\mathrm{kHz}$ and $(\OmegaX+\OmegaY)/4\pi = [116 + (246-116)x^2]\,\mathrm{kHz}$ where $x = (\supp - 0.7)/(0.07 - 0.7)$. In the full graph, the frequency splitting $\delta$ is determined by $\supp$ according to $\supp = 2 (\gX^2+\gY^2)/\kappa \delta$.}
    \label{Fig3}
\end{figure}

In the weak coupling, resolved sidebands regime, the optically induced width of a mechanical mode is $ 4 g^2/\kappa$. We can quantify the spectral overlap between the $X$ and $Y$ modes using the ratio of their frequency splitting $\delta = (\OmegaX-\OmegaY)$ to their mean width, defining an overlap parameter as $\supp = 2 (\gX^2+\gY^2)/\kappa \delta$. 
For $\supp \ll 1 $
the two-dimensional system can be approximately described as the combination of two independent oscillators, while for  $\supp \ge 1$ 
the full two-dimensional dynamics emerges, and the correlation between the two orthogonal directions becomes significant. To summarize the information on both the state purity and the spectral overlap, we show in Fig. \ref{Fig3} the plot of the $(\mu, \supp)$  
parameter space, where we compare our system with the previous results reported in the literature. 

The relevance of the correlation between the $X$ and $Y$ projections of the two-dimensional motion can be quantified by the quantum discord $\mathcal{D}_{X \leftarrow Y}$ ($\mathcal{D}_{Y \leftarrow X}$), defined as the difference between the mutual information between the two subsystems, and the one-way classical correlations. The latter is the maximum amount of information obtainable on $X$ ($Y$) by locally measuring the sub-system $Y$ ($X$). A positive discord, even on separable (not entangled) states, is an indicator of quantumness \cite{Ollivier2002}. For bipartite Gaussian states, the expression of the quantum discord can be written in a close form using the four symplectic invariants of the covariance matrix \cite{Giorda2010,Adesso2010,Olivares2012}. We obtain $\mathcal{D}_{X \leftarrow Y} = 0.0423\pm0.0007$ and $\mathcal{D}_{Y \leftarrow X} = 0.0471\pm0.0012$. A contour plot of the symmetrized discord $0.5 (\mathcal{D}_{X \leftarrow Y} + \mathcal{D}_{Y \leftarrow X})$ is displayed in Fig. \ref{Fig3} showing that, as expected, a larger quantum discord is achieved at increasing spectral overlap. Even if a small but positive discord can exist even for states with low purity, and it is indeed a rather general feature \cite{Ferraro2010}, in our case the discord constitutes a sizeable part of the total mutual information, roughly half of it, so quantum correlations are of the same order as total correlations. 

The description based on the original $X$ and $Y$ modes does not capture the full physical properties of the optomechanical system. As $\splitt$ decreases, the motion is better understood using a description based on the geometric bright and dark modes, corresponding to directions parallel and orthogonal to the cavity axis, respectively. With this basis, it has been shown that the two-dimensional cooling becomes less effective since the dark mode is not directly coupled to the optical field. Moreover, as the strong optomechanical coupling is approached, optical cooling becomes less efficient as it assumes a sublinear dependence on $g^2$. As an additional effect of strong coupling, the identification of two orthogonal oscillation directions as approximate eigenvectors of the complete optomechanical system becomes poorly accurate~\cite{Ranfagni2021}. We derive two considerations. 
The first is that, for a fair description of the system, we need to abandon the ($X-Y$) coordinate system, and prioritize indicators that are independent of any specific reference frame. The global state purity already satisfies this requirement. For quantum discord, we evaluated the maximum of $\,\mathcal{D}_{\phi \,\leftarrow (\phi+\pi/2)} $ where $\phi$ defines the projection of the two-dimensional motion along the $\phi$ direction. We obtain a value of $0.0482\pm0.0012$ for an angle $\phi = -9^{\circ}$.
The second consideration is that simultaneously achieving a low effective temperature (i.e., high state purity) and large spectral overlap (i.e., strongly two-dimensional characteristics) is not obvious. Appropriate tuning of the system parameters allows one to maximize either the purity or the discord, and the quantum indicators pair can thus be optimized for specific applications.

\section*{Conclusions}
 
In conclusion, we have achieved two-dimensional motion of a nanosphere in an optical potential where not only the oscillations predominantly occupy the quantum ground state in all directions across the plane, setting a new benchmark for the purity of the two-dimensional state, but also significant correlations are consistently present between any pair of orthogonal directions.
Therefore, the system behavior cannot be reduced to a simple decomposition into two one-dimensional modes. Instead, the motion exhibits distinct two-dimensional characteristics that can be detected spectrally.

The measured correlations are not yet strong enough to produce entanglement between mechanical modes (i.e., between oscillations along two directions of the plane). In fact, it has been shown theoretically that achieving this type of entanglement is almost impossible with our setup in its present configuration, where the background is at room temperature and only a single mode of the electromagnetic field is present \cite{Vitali2007,Genes2008}. However, the system we have developed, characterized by high purity and strong spectral correlations, represents an excellent platform for achieving entanglement, for example by introducing additional electromagnetic fields \cite{Hartmann2008,Li2015}. By adding blue-detuned laser fields to the tweezer light, with a controlled phase relationship with respect to the cooling radiation, one could implement schemes similar to those successfully realized in ultra-cryogenic microwave experiments, where entanglement between oscillators quadratures is achieved \cite{Ockeloen-Korppi2018,Kotler2021,Mercier2021}. Furthermore, as proposed in Ref. \cite{Mazzola2011}, an additional cavity would allow to activate entanglement between the two mechanical oscillators exhibiting quantum discord. Pulsed schemes where the entangling fields enter the cavity through its output port are also promising \cite{Rakhubovsky2020}.
The realization of mechanical entanglement would mark a significant milestone in the development of innovative quantum information schemes \cite{Weedbrook2012}, as well as for the study of quantum decoherence at the macroscopic level \cite{Marshall2003,Bassi2013,Gasbarri2021}.

\section*{Acknowledgments}
We acknowledge financial support from PNRR MUR Project No. PE0000023-NQSTI and by the European Commission-EU under the Infrastructure I-PHOQS “Integrated Infrastructure Initiative in Photonic and Quantum Sciences ” [IR0000016, ID D2B8D520, CUP D2B8D520].

\bibliography{database}


\end{document}


\title[]{Supplementary Material for: High purity two-dimensional levitated mechanical oscillator}

\author{Q. Deplano}
\affiliation{Dipartimento di Fisica e Astronomia, Universit\`a di Firenze, via Sansone 1, I-50019 Sesto Fiorentino (FI), Italy}
\affiliation{INFN, Sezione di Firenze, via Sansone 1, I-50019 Sesto Fiorentino (FI), Italy}

\author{A. Pontin}
\affiliation{CNR-INO, largo Enrico Fermi 6, I-50125 Firenze, Italy}

\author{A. Ranfagni}
\affiliation{Dipartimento di Fisica e Astronomia, Universit\`a di Firenze, via Sansone 1, I-50019 Sesto Fiorentino (FI), Italy}

\author{F. Marino}
\affiliation{CNR-INO, largo Enrico Fermi 6, I-50125 Firenze, Italy}
\affiliation{INFN, Sezione di Firenze, via Sansone 1, I-50019 Sesto Fiorentino (FI), Italy}

\author{F. Marin}
\email[Electronic mail: ]{francesco.marin@unifi.it}
\affiliation{Dipartimento di Fisica e Astronomia, Universit\`a di Firenze, via Sansone 1, I-50019 Sesto Fiorentino (FI), Italy}
\affiliation{European Laboratory for Non-Linear Spectroscopy (LENS), via Carrara 1, I-50019 Sesto Fiorentino (FI), Italy}
\affiliation{INFN, Sezione di Firenze, via Sansone 1, I-50019 Sesto Fiorentino (FI), Italy}
\affiliation{CNR-INO, largo Enrico Fermi 6, I-50125 Firenze, Italy}


\maketitle

\tableofcontents

\section*{Model}

The motion in the transverse tweezer plane is described in terms of dimensionless position and momentum operators $\xX, \pX, \xY, \pY$ obtained by normalizing the corresponding physical variables to their respective zero-point fluctuations $\xzpf = \sqrt{\hbar/2 m \OmegaXY}$ and $\pzpf = \sqrt{\hbar m \OmegaXY/2 }$ ($m$ is the mass of the nanosphere, $\OmegaXY$ the oscillation frequencies of the $X$ and $Y$ modes in the absence of optomechanical interaction). The Langevin equations of motion are:
\begin{eqnarray}
\dot{x} & = & \OmegaX \,\pX 
\label{eq_dxdt}\\
\dot{p}_{\scriptstyle{x}} &=& -\OmegaX \,\xX -\gamma_x \,\pX + 2 \gX\, Q + 2 \sqrt{\GX} \,\xiX \\
\dot{y} & = & \OmegaY \,\pY \\
\dot{p}_{\scriptstyle{y}} &=& -\OmegaY \,\xY -\gamma_y \,\pY + 2 \gY \,Q + 2 \sqrt{\GY} \,\xiY \\
\dot{Q} &=& -\Delta P -\frac{\kappa}{2} \,Q + \sqrt{\kappa}\, Q_{\mathrm{in}}  \\
\dot{P} &=& \Delta Q -\frac{\kappa}{2}\, P + 2 \gX \,\xX + 2 \gY\, \xY + \sqrt{\kappa}\, P_{\mathrm{in}}
\label{eq_dPdt}
\end{eqnarray}
where $\,Q\,=\,(a\,+\,a^{\dag})\,$ and  $\,P\,=\,i(a^{\dag}\,-\,a)\,$ are the quadratures of the intracavity field $a$, and $\,Q_{\mathrm{in}}\,=\,(a_{\mathrm{in}}\,+\,a_{\mathrm{in}}^{\dag})\,$,  $\,P_{\mathrm{in}}\,=\,i(a_{\mathrm{in}}^{\dag}\,-\,a_{\mathrm{in}})\,$ are the quadratures of the input vacuum noise. The noise sources have non-null correlation functions $\langle \xiX(t) \, \xiX(t')\rangle = \langle \xiY(t) \, \xiY(t')\rangle = \langle a_{\mathrm{in}}(t) \, a^{\dag}_{\mathrm{in}}(t')\rangle= \delta(t-t')$ (due to the high background temperature, we can classically model the mechanical noise). 

The gas damping rates $\gamma_j$ play a negligible role at low pressure, where optomechanical damping largely dominates, and we fix them at the nominal value of $10^{-4} \,\mathrm{s}^{-1}$. The decoherence of the motion is mainly due to collisions with the background gas, and to quantum noise in the dipole scattered light. The corresponding rates $\GXY$ are left as free parameters in the analysis of the experimental data, and the inferred values well agree, within their uncertainties, with their theoretical estimates \cite{Ranfagni2022}. In particular, for this experiment such agreement is obtained by assuming a background pressure in the range $(2 \div 3) \times 10^{-8}\,$mbar, which is indeed the indication given by two pressure gauges in different positions of the vacuum chamber.  
The optomechanical coupling rates $\gX$ and $\gY$ are proportional to $\sin^2 \theta$ and $\sin \theta \cos \theta$ respectively. Also in this case, we consider them as free parameters in the data analysis, the agreement with the expected values is good \cite{Ranfagni2022}, and from their ratio we deduce the actual angle of the polarization axis. 

From the Langevin equations (\ref{eq_dxdt} - \ref{eq_dPdt}) we derive the drift and diffusion matrices used in the main text for the Lyapunov equation.

The Langevin equations can be written in the Fourier space as
\begin{eqnarray}
\tQo &=& i \ccm \left[g_x \txo+g_y \tyo  \right]+\tilde{Q}_\mathrm{in} \left(\omega\right)
\label{eq_Lang1} \\
\txo &=&  2  \gX \cxm \tQo + \tNx 
\label{eq_Lang2} \\
\tyo &=& 2  \gY \cym \tQo + \tNy 
\label{eq_Lang3}
\end{eqnarray}
where the tilde denotes Fourier transformed (FT) operators $\tilde{O}\left(\omega\right)=\mathrm{FT}\left[\hat{O}\left(t\right)\right]$, and
$\tilde{O}^{\dagger}\left(\omega\right)=\mathrm{FT}\left[\hat{O}^{\dagger}\left(t\right)\right]$.
We have defined the susceptibilities
\begin{eqnarray}
\chi_j\left(\omega\right) &=& \frac{\Omega_j}{\Omega_j^2-\omega^2-i\omega\gamma_j}\\
\chi_\mathrm{c}\left(\omega\right) &=& \frac{1}{-i
\left(\Delta+\omega\right)+\kappa/2}
\end{eqnarray}
with $j= (x, y)$, and the compact form
$\,\ccm=\chi_\mathrm{c}\left(\omega\right) \,-\,
\chi_\mathrm{c}^{*}\left(-\omega\right)$.
The noise input terms can be written as
\begin{gather}
\tilde{Q}_\mathrm{in} \left(\omega\right) = \sqrt{\kappa}\left[\chi_\mathrm{c} \om \tilde{a}_{\mathrm{in}}\om +
\chi_\mathrm{c}^{*} \left(- \omega\right) \tilde{a}^{\dagger}_{\mathrm{in}}\left(  \omega \right)  \right]\\
\tNi=2 \sqrt{\Gamma_i} \chi_i\om \tilde{\xi}_i   \, .
\end{gather}

We define the bright mode as  $x_\mathrm{b} = \frac{1}{g_\mathrm{b}}\left(\gX x+ \gY y\right)$. Using it in Eq. (\ref{eq_Lang1}) and combining Eqs. (\ref{eq_Lang2}-\ref{eq_Lang3}) we obtain two equations respectively for $\tilde{Q}$ and $\tilde{x}_{\mathrm{b}}$. Inserting the first into the second we derive 
\begin{equation}
   \tilde{x}_\mathrm{b} \left( \omega \right)=\frac{2\left(\gX^2\chi_x +\gY^2 \chi_y \right) \tilde{Q}_\mathrm{in}  \left(\omega\right) +\gX \tNx + \gY \tNy}{g_\mathrm{b}\left[1-2 i \ccm \left( \gX^2 \chi_x + \gY^2 \chi_y \right) \right]}
\end{equation}
from which we calculate the displacement noise spectrum
\begin{equation}
S_{x_\mathrm{b} x_\mathrm{b}}\om=\frac{4}{\gb^2}\,
\frac{\gX^2 \,\GX \left|\chi_x\right|^2
+\gY^2 \,\GY\left|\chi_y\right|^2+\left|\gX^2 \chi_x+\gY^2 \chi_y\right|^2
\kappa \left|\chi_\mathrm{c}\left(-\omega\right)\right|^2}
{\left|1-2 i \chi_\mathrm{c}^{-}\left(\gX^2 \chi_x+\gY^2 \chi_y\right)\right|^2}
\end{equation}
used in the main text.

The cavity output field is given by the input-output relation 
$a_{\mathrm{out}} = \sqrt{\kappa} a-a_{\mathrm{in}}$, and
the heterodyne spectrum normalized to shot noise can be written as
\begin{equation}
S_{\mathrm{out}}(\Omega_{\mathrm{LO}}+\omega) 
= 1 + \eta  \,\kappa \,\left|\chi_\mathrm{c}\left(\omega\right)\right|^2 \,\gb^2 \, S_{x_\mathrm{b} x_\mathrm{b}}\om  \, .
\end{equation}

The parameter $g_{\mathrm{b}}$ plays no meaningful role in the model, and indeed it cancels out in the expression of the heterodyne spectrum. We note however that, by defining it as 
\begin{gather}
g_\mathrm{b}=\sqrt{\frac{\gX^2\Omega_x+\gY ^2 \Omega_y}{\Omega_{\mathrm{b}}}}
\end{gather}
where
\begin{gather}
\Omega_{\mathrm{b}}=\sqrt{\frac{\gX^2 \Omega_{x}^3+\gY^2 \Omega_{y}^3}{\gX^2\Omega_{x}+\gY^2 \Omega_{y}}} 
\label{eq_Omegab}
\end{gather}
and assuming that $\gX \propto \sin^2 \theta$ and $\gY \propto \sin \theta \cos \theta$, $x_{\mathrm{b}}$ can be identified as the motion along the cavity axis, normalized to the zero-point fluctuations $\sqrt{\hbar/2 m \Omega_{\mathrm{b}}}$.

\subsection*{Ground state probability}

If the correlations between the two mechanical modes are negligible, the probability of the two-dimensional ground state is simply given by $1/(\nX+1)(\nY+1)$, and it is larger than 0.25 whenever both thermal occupancies are below unity. In the general case this expression is not valid, but the ground state probability can be calculated from the covariance matrix, with an integration on the phase-space. Defining the vector of phase-space variables $R=(x,\, p_x,\, y,\, p_y)$, the Wigner function of the two-dimensional Gaussian state is \cite{Paris2003}
\begin{equation}
    W(R) = \frac{1}{(2\pi)^2\,\sqrt{\mathrm{Det}[\Vm]}}\,\exp\left(-\frac{1}{2} R (\Vm)^{-1} R^{\mathrm{T}}\right)
\end{equation}
and the projector on the two-dimensional ground state, expressed in the phase-space, is 
\begin{equation}
g(R) = 4 \exp \left(-\frac{1}{2} R\,R^{\mathrm{T}}\right)  \, .
\end{equation}
Finally, the probability that the two-dimensional oscillator is in its ground state is given by
\begin{equation}
    \mathrm{P}(0,0) = \int W(R)\,g(R)\, \mathrm{d}^4R  \,.
\end{equation}
The calculated ground state probability for our system is reported in Table \ref{Tab_Suppl_2}.

\subsection*{Quantum discord of a bipartite Gaussian state}

The quantum discord is calculated for a two-mode squeezed thermal state in Ref.
 \cite{Giorda2010}, 
and for a general bipartite Gaussian state in Ref.
\cite{Adesso2010}. 
A simplified expression is given in Ref. 
\cite{Olivares2012}. 
For the sake of completeness, we report below its full evaluation starting from our definition of phase-space operators.

The covariance matrix $\Vm$ can be written using 2x2 blocks {\boldmath $\alpha$}, {\boldmath $\beta$} and {\boldmath $\gamma$} in the form
\begin{displaymath}
    \Vm_{\scriptstyle{\phi}} = 
    \left( \begin{array}{cc}
    \mbox{\boldmath $\alpha$} & \mbox{\boldmath $\gamma$} \\
    \mbox{\boldmath $\gamma$}^{\scriptstyle{T}}  &  \mbox{\boldmath $\beta$}
    \end{array}  \right)
\end{displaymath}
The correlations between the orthogonal modes are contained in the off-diagonal  block {\boldmath $\gamma$}.

Four symplectic invariants are defined as $I_1=\mathrm{Det}[\mbox{\boldmath $\alpha$}]$, $I_2=\mathrm{Det}[\mbox{\boldmath $\beta$}]$, $I_3=\mathrm{Det}[\mbox{\boldmath $\gamma$}]$, and $I_4=\mathrm{Det}[\Vm]$. With our definitions, describing a physical system requires $\,I_1, I_2 \ge 1\,$ and $\,d_{\pm} \ge 1$, where the symplectic eigenvalues are defined by 
$\,d_{\pm} = \sqrt{(D \pm \sqrt{D^2-4 I_4} ) /2}\,$ 
with $D = I_1+I_2+2 I_3$. 

If $\,(I_4- I_1 I_2)^2\le (1+I_2) I_3^2 (I_1+I_4)\,$ (a condition which is verified in our case), the quantum discord can be calculated as
\begin{equation}
    \mathcal{D}_{X \leftarrow Y}\,=\,f\left(\sqrt{I_2}\right)-f\left(d_{+}\right)-f\left(d_{-}\right)+f\left(E\right)
\end{equation}
where 
\begin{equation}
    f(x) = \frac{x+1}{2} \,\log{\left(\frac{x+1}{2}\right)}-\frac{x-1}{2} \,\log{\left(\frac{x-1}{2}\right)}
\end{equation}
($\log$ is the natural logarithm) and
\begin{equation}
    E\,=\,\frac{|I_3| + \sqrt{I_3^2+\left(I_2-1\right)\left(I_4-I_1\right)}}{I_2-1}   \, .
\end{equation}
The expression of $\mathcal{D}_{Y \leftarrow X}$ is obtained from that of $\mathcal{D}_{X \leftarrow Y}$ by swapping $I_1$ and $I_2$.

\subsection*{Quantum discord in a rotated frame  }

To define the phase-space in a geometrically rotated frame, we must consider that the modal eigenfrequency depends on the direction (see Eq.
\ref{eq_Omegab}) 
and therefore also the zero-point fluctuations used to normalize the position and momentum variables. However, the symplectic invariants of the covariance matrix do not depend on this normalization.
We then removed the normalization to the zero-point fluctuations of the $X$ and $Y$ modes, avoiding to rely on their uncoupled resonance frequencies $\OmegaXY$, and computed the covariance matrix of the mechanical system in an arbitrary reference frame as $\Vm_{\scriptstyle{\phi}} = R(\phi) N \Vm N R^{\scriptstyle{T}}(\phi)$. Here $R(\phi)$ rotates by an angle $\phi$ around the $Z$ axis, and $N=\mathrm{Diag}[1/\sqrt{\Omega_x}, \sqrt{\Omega_x}, 1/\sqrt{\Omega_y}, \sqrt{\Omega_y}]$. The quantum discord $\,\mathcal{D}_{\phi \,\leftarrow (\phi+\pi/2)} $ is calculated from $\Vm_{\scriptstyle{\phi}}$. 

\section*{Additional data set}

Thanks to a pair of ring electrodes positioned on the tweezer optics holder at about $1\,$mm from the nanosphere, we can generate a static electric field along the tweezer axis. Since the nanosphere is generally charged \cite{Deplano2024}, this allows us to displace its equilibrium position. In this way, we slightly tune the oscillation frequencies and, compensating for the tweezer radiation pressure, maximize the frequency splitting between $X$ and $Y$. 

We acquired two additional data sets, at different applied voltages. Unfortunately, in these data sets the signals given by the rotational motion are more evident, and we had to post-select the time series in order to delete the time intervals in which the erratic rotational peaks cross the spectral regions of interest. The overall signal-to-noise ratio is therefore lower than in the spectra described in the main text. 

\begin{figure}[!htb]
    \centering
      \includegraphics[scale=0.65]{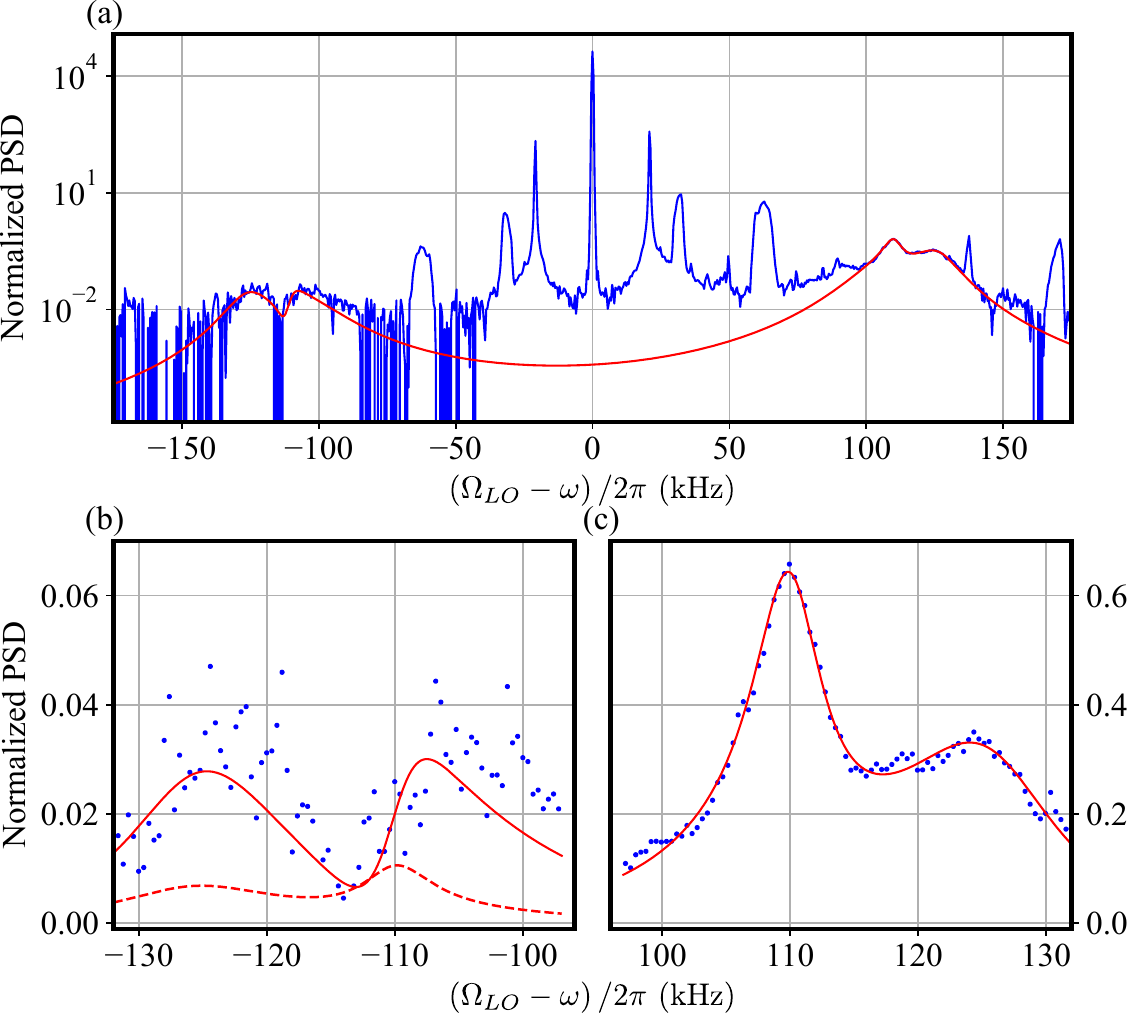}
    \caption[Fig2]{\textbf{Power Spectral Density of the heterodyne signal for an additional data set.} The spectrum is normalized to the measured shot noise which is then subtracted (the dark noise is preliminarily subtracted from all spectra). The abscissa is the frequency difference with respect to the local oscillator. The detuning is $\Delta/2\pi = -110\,$kHz, and a voltage of $35\,$V is applied to the electrodes on the tweezer. The red solid line shows the fit of Eqs. (\ref{eq_Sout}-\ref{eq:Sxb}) to the experimental right sideband.  The lower panels display enlarged views of the left (b) and right (c) sidebands. The dashed red line in panel (b) is obtained by replacing $S_{x_\mathrm{b} x_\mathrm{b}}\om\,\to\,S_{x_\mathrm{b} x_\mathrm{b}}\left( -\omega \right)$.}
    \label{Fig_Suppl}
\end{figure}

An example is shown in Fig. \ref{Fig_Suppl}. In panel (b) of the figure, which displays an enlarged view of the left sideband, we also report, together with the theoretical curve fitted to the right sideband, the same theoretical signal with $S_{x_\mathrm{b} x_\mathrm{b}}\om \to S_{x_\mathrm{b} x_\mathrm{b}}\left( -\omega \right)$. It shows how the heterodyne signal would appear in the case of a mechanical system dominated by classical noise, i.e., when the displacement noise spectrum is frequency symmetric and the difference between the two motional sidebands is due exclusively to the cavity filtering. The huge difference, both in amplitude and shape, compared to the actual spectral signal is further evidence of the highly quantum nature of the two-dimensional oscillator.   

In Table \ref{Tab_Suppl_1} we summarize the fitted parameters for all data sets (including the first one, discussed in the main text), and in Table \ref{Tab_Suppl_2} we report the extracted values of several physical quantities characterizing the two-dimensional system.

\begin{table}
\begin{tabular}{c||c|c|c}
Voltage  (V)  &  0 & 22.5 & 35\\
  $\OmegaX/2\pi$ (Hz)  &  $122170\pm120$ & $122290\pm280$ & $121610\pm160$ \\
   $\OmegaY/2\pi$ (Hz)   &  $109370\pm150$ & $108970\pm220$ & $107640\pm150$ \\
 $\gX/2\pi$ (Hz)    & $14130\pm220$ & $14420\pm230$ & $15160\pm310$ \\
   $\gY/2\pi$ (Hz)   &  $10370\pm160$ & $10300\pm190$ & $10060\pm 110$ \\
 $\GX/2\pi$ (Hz)    &  $4030\pm200\pm120$ & $3890\pm220\pm150$ & $3250\pm180\pm100$ \\
  $\GY/2\pi$ (Hz)    & $3050\pm170\pm90$ & $2990\pm160\pm70$ & $2520\pm140\pm70$ \\
\end{tabular}
\caption{Parameters extracted from the fit of three data sets with the model of Eqs. (\ref{eq_Sout}-\ref{eq:Sxb}). The reported uncertainty is the spread (one standard deviation) on five independent acquisitions. For the decoherence rate, the second quoted error derives from a spread of $5\%$ in the detection efficiency, which is a conservative estimate of the uncertainty of an independent measurement yielding $\eta = 0.32$.}
\label{Tab_Suppl_1}
\end{table}

\begin{table}
\begin{tabular}{c||c|c|c}
Voltage  (V)  &  0 & 22.5 & 35\\
  $\nX$  &  $0.551\pm0.003\pm0.027$ & $0.514\pm0.001\pm0.026$ & $0.407\pm0.002\pm0.020$ \\
   $\nY$    &  $0.740\pm0.013\pm0.036$ & $0.716\pm0.007\pm0.036$& $0.626\pm0.007\pm0.031$  \\
 purity   & $0.209\pm0.002\pm0.011$ & $0.219\pm0.002\pm0.011$ & $0.260\pm0.002\pm0.012$ \\
 $\mathcal{D}_{X \leftarrow Y}$ & $0.0423\pm0.0003\pm0.0004$ & $0.0361\pm0.0010\pm0.0006$ & $0.0415\pm0.0011\pm0.0008$ \\
 $\mathcal{D}_{Y \leftarrow X}$ & $0.0471\pm0.0006\pm0.0006$ & $0.0399\pm0.0018\pm0.0004$  & $0.0449\pm0.0016\pm0.0007$  \\
 P(0,0) & $0.386\pm0.003\pm0.014$ & $0.449\pm0.002\pm0.014$ & $0.399\pm0.003\pm0.014$
\end{tabular}
\caption{Thermal occupancies of the $X$ and $Y$ modes considered as one-dimensional oscillators, two-dimensional state purity, quantum discord, and probability of the two-dimensional ground state P(0,0), deduced for three different data sets. The first reported uncertainty is the statistical spread over five independent acquisitions (one standard deviation), the second quoted error derives from an uncertainty of $5\%$ in the detection efficiency.}
\label{Tab_Suppl_2}
\end{table}

\bibliography{database}